\documentclass[11pt]{article}
\usepackage{amsfonts}
\usepackage{amssymb}
\usepackage{amsmath}
\usepackage{graphicx}

\def \Zed {\mbox{$\mathbb{Z}$}}

\def \Natural {\mbox{$\mathbb{N}$}}

\def \lket {\left|}
\def \rket {\right\rangle}

\newcommand{\ket}[1]{\lket #1\rket}

\def \cent
        {\mbox{c\hspace{-0.85mm}\rule[-0.4mm]{0.05mm}{2.4mm}}\hspace{0.85mm}}
\def \qed {\hfill \rule{0.2cm}{0.2cm}\vspace{3mm}}
\newcommand{\inpar}[3]{
        \setlength{\twidth}{\textwidth}
        \addtolength{\twidth}{-#1}
        \rule{#1}{0mm}\parbox{\twidth}{#3}\\[#2]}
\newenvironment{mylist}[1]
        {\begin{list}{}{\setlength{\leftmargin}{#1}
        \setlength{\rightmargin}{0.0cm}\setlength{\labelsep}{1.3mm}
        \setlength{\labelwidth}{0.8cm}\setlength{\itemsep}{0.0cm}}}
        {\end{list}}
\newtheorem{theorem}{Theorem}
\newtheorem{lemma}[theorem]{Lemma}

\begin{document}
\newlength{\twidth}

\title{\Large\bf Two-way finite automata with quantum and classical states}

\author{
Andris Ambainis\\
Computer Science Division\\
University of California\\
Berkeley, CA 94720\\
ambainis@cs.berkeley.edu
\and
John Watrous\\
Department of Computer Sciences\\
University of Calgary\\
Calgary, Alberta, Canada T2N 1N4\\
jwatrous@cpsc.ucalgary.ca}

\maketitle

\thispagestyle{empty}

\begin{abstract}
We introduce 2-way finite automata with quantum and classical states (2qcfa's).
This is a variant on the 2-way quantum finite automata (2qfa) model
which may be simpler to implement than unrestricted 2qfa's; the internal state
of a 2qcfa may include a quantum part that may be in a (mixed) quantum state,
but the tape head position is required to be classical.

We show two languages for which 2qcfa's are better than classical 
2-way automata. First, 2qcfa's can recognize palindromes, a
language that cannot be recognized by 2-way deterministic or
probabilistic finite automata. Second, in polynomial time 2qcfa's
can recognize $\{ a^n b^n |n\in\Natural\}$, a language that
can be recognized classically by a 2-way probabilistic automaton but
only in exponential time.
\end{abstract}


\section{Introduction}
\label{sec:introduction}

The theory of quantum computing has had some remarkable successes, such as
Shor's quantum algorithm for factoring integers in polynomial
time \cite{Shor97} and Grover's algorithm for searching an unordered list 
of size $n$ with just $O(\sqrt{n})$ accesses to the list \cite{Grover96}. 
However, these algorithms are for general quantum Turing 
machines or quantum circuits.
Today's experimental quantum computers are much less
powerful and the biggest quantum computer consists of
just 3 quantum bits (qubits).
Therefore, it may be interesting to consider more 
restricted theoretical models of quantum computers. 
In this paper, we consider the following question:
what is the simplest and most restricted model of
computation where quantum computers are still more
powerful than their classical counterparts?
Classically, one of simplest models of computation
is a finite automaton. Quantum finite automata
have been recently studied by several authors \cite{AmbainisF98,
AmbainisN+99,BrodskyP99,KondacsW97,MooreC97}.

Two models of quantum finite automata have been considered.
The simplest is 1-way quantum finite automata (1qfa's) introduced by
\cite{KondacsW97,MooreC97}.
This is very simple model of computation but it is
not very powerful; the languages recognized by 1qfa's
form a proper subset of the regular languages (languages
recognized by 1-way deterministic automata).
A more powerful generalization of this model is 1-way qfa's that allow mixed
states (defined similarly to quantum circuits with mixed
states \cite{AharonovK+98}).
Any 1-way dfa can be easily simulated by a 1-way qfa with mixed states.
However, all languages recognized by 1-way qfa's with mixed states (with
bounded error) are still regular.
The second model is 2-way quantum finite automata (2qfa's) \cite{KondacsW97}.
In this model, it is easy
to simulate any deterministic automaton and some non-regular
languages can be recognized as well. This implies that
2qfa's are strictly more powerful than their classical 
counterparts. However, this model has another disadvantage:
it allows superpositions where the head can be in multiple
positions simultaneously.
To implement such a machine, we need at least $O(\log n)$ qubits
to store the position of the head (where $n$ is the length
of the input). It would be nicer to have a model where the
size of the quantum part does not depend on the length
of the input.

In this paper, we propose 2-way finite automata with quantum and classical
states (2qcfa's), an intermediate model between
1qfa's and 2qfa's. This model is both powerful (2qcfa's
can trivially simulate any classical automaton and recognize some
languages that classical automata cannot) and can be implemented
with a quantum part of constant size. 

We consider the following two languages:
$L_{pal}=\{x\in\{a,b\}^{\ast}\,|\,x = x^R\}$ (the language consisting of all
palindromes over the alphabet $\{a,b\}$) and
$L_{eq}=\{a^nb^n\,|\,n\in\Natural\}$.
It has been shown that no probabilistic 2-way finite automaton can recognize
$L_{pal}$ with bounded error in any amount of time~\cite{DworkS92}, and that
no classical 2-way finite automaton can recognize $L_{eq}$ (or any other
nonregular language) with bounded error in polynomial
time~\cite{DworkS90, KanepsF91}.
We prove that there exists an exponential time 2qcfa recognizing $L_{pal}$
with bounded probability of error, and a polynomial time 2qcfa recognizing
$L_{eq}$ with bounded probability of error, thereby giving two examples
where 2qcfa's are provably more powerful than classical 2pfa's.

It is interesting to note that our 2qcfa's for $L_{pal}$ and $L_{eq}$ require
that the quantum part of the machine consist of only a single qubit; in
essence, our 2qcfa's use the quantum state of this qubit to represent and
process certain information regarding the input.
While the extremely high precision required in manipulating this single qubit
certainly calls into question the practicalities of these algorithms, it is
interesting that such extreme examples of space-efficiency/precision trade-offs
exist, particularly in light of existing bounds on the amount of
information transmittable and accessible in a single qubit (or finite
collection of qubits)~\cite{AmbainisN+99,Holevo73,Nayak99}.

The remainder of this paper has the following organization.
In Section~\ref{sec:definitions} we give a definition of 2-way finite
automata with quantum and classical states.
In Section~\ref{sec:palindromes} we describe a 2qcfa for $L_{pal}$
and in Section~\ref{sec:anbn} we give a 2qcfa for $L_{eq}$.
We conclude with Section~\ref{sec:conclusion}, which includes mention of
various open questions relating to this paper.


\section{Definitions}
\label{sec:definitions}

In this section we give our definition for 2-way finite automata with
quantum and classical states.
Informally, we may describe a 2qcfa as a classical 2-way finite automaton that
has access to a fixed-size quantum register, upon which it may perform quantum
transformations and measurements.
The transformations and measurements are determined by local descriptions of
the classical portion of the machine, and the results of the measurements may
determine the manner in which the classical part of the machine evolves.

Before giving a more formal definition of our model, we recall a few basic
facts regarding quantum computing.
For a more detailed overview of quantum computing, we refer the reader
to~\cite{Berthiaume97,Kitaev97,Preskill98}.
Let $Q$ be a finite set.
A {\em superposition} of elements in $Q$ is a norm 1 vector in a Hilbert space
$\mathcal{H}$ of dimension $|Q|$, where each element $q\in Q$ is identified
with an elementary unit vector denoted by $\ket{q}$.
Any superposition may therefore be written in the form
$\sum_{q\in Q}\alpha_q\ket{q}$, where each $\alpha_q$ is a complex number and
we have $\sum_{q\in Q}|\alpha_q|^2 = 1$.
In general we denote superpositions as $\ket{\phi}$, $\ket{\psi}$, etc., even
when symbols $\phi$, $\psi$, etc., are not used alone.
A unitary operator on $\mathcal{H}$ is any invertible linear operator that
preserves length (equivalently $U$ is unitary if $U^{-1} = U^{\dagger}$, where
$U^{\dagger}$ denotes the adjoint of $U$).
When we say that we apply the {\em unitary transformation} described by $U$ to
a system in a given superposition $\ket{\psi}$, we mean that the superposition
of this system is changed according to the mapping
$\ket{\psi}\mapsto U\ket{\psi}$.
A set of operators $\{P_j\}$ on $\mathcal{H}$ specify an {\em orthogonal
measurement} (also called a {\em von~Neumann measurement}) if
$P_j=P_j^{\dagger}$ and $P_j^2=P_j$ for all $j$, $P_j P_k=0$ for $j\not=k$,
and $\sum_j P_j = I$.
If a superposition $\ket{\psi}$ is measured (or observed) via a measurement
described by a collection $\{P_j\}$, the following happens: (i) the result of
the measurement is $j$ with probability $\|P_j\ket{\psi}\|^2$ for each $j$,
and (ii) the superposition of the system is changed to
$\frac{1}{\|P_j\ket{\psi}\|}P_j\ket{\psi}$ for whichever $j$ was the result of
the measurement.

Now we may define 2qcfa's more precisely.
A two-way finite automaton with quantum and classical states is specified by a
9-tuple
\[
M = (Q,S,\Sigma,\Theta,\delta,q_{0},s_{0},S_{acc},S_{rej}),
\]
where $Q$ and $S$ are finite state sets (quantum states and classical states,
respectively), $\Sigma$ is a finite alphabet, $\Theta$ and $\delta$ are
functions described below that specify the behavior of $M$, $q_{0}\in Q$ is
the initial quantum state, $s_{0}\in S$ is the initial classical state, and
$S_{acc},S_{rej}\subseteq S$ are the sets of (classical) accepting states and
rejecting states, respectively.
We let $\Gamma = \Sigma\cup\{\cent,\$\}$ be the tape alphabet of $M$, where
$\cent\not\in\Sigma$ is called the left end-marker and $\$\not\in\Sigma$ is
called the right end-marker.

The function $\Theta$ specifies the evolution of the quantum portion of the
internal state: for each pair
$(s,\sigma)\in S\backslash(S_{acc}\cup S_{rej})\times\Gamma$,
$\Theta(s,\sigma)$ is an action to be performed on the quantum portion of the
internal state of $M$.
Each action $\Theta(s,\sigma)$ corresponds to either a unitary transformation
or an orthogonal measurement.

The function $\delta$ specifies the evolution of the classical part of $M$
(i.e., the classical part of the internal state and the tape head).
In case $\Theta(s,\sigma)$ is a unitary transformation, $\delta(s,\sigma)$ is
an element of $S\times\{-1,0,1\}$ specifying a new classical state and a
movement of the tape head.
In case $\Theta(s,\sigma)$ is a measurement, $\delta(s,\sigma)$ is a mapping
from the set of possible results of the measurement to $S\times\{-1,0,1\}$
(again specifying a new classical state and a tape head movement, this time
one such pair for each outcome of the observation).
It is assumed that $\delta$ is defined so that the tape head never moves
left when scanning the left end-marker \cent, and never moves right when
scanning the right end-marker \$.

On a given input $x$, a 2qcfa $M$ is to operate as follows.
Initially, the classical part of $M$'s internal state is in state $s_0$, the
quantum part of the internal state is in superposition $\ket{q_0}$, and the
tape head of $M$ is scanning the tape square indexed by 0.
The tape squares indexed by $1,\ldots,|x|=n$ contain $x_1,\ldots,x_n$, while
the squares indexed by $0$ and $n+1$ contain end-markers \cent\ and \$,
respectively.
On each step, the quantum part of the internal state is first changed
according to $\Theta(s,\sigma)$, where $s$ is the current classical internal
state and $\sigma$ is the currently scanned tape symbol, and then the
classical internal state and tape head position are changed according to
$\delta(s,\sigma)$ (along with the particular result obtained from
$\Theta(s,\sigma)$ in case $\Theta(s,\sigma)$ is a measurement).

Since the results obtained from each measurement $\Theta(s,\sigma)$ are
probabilistic, the transitions among the classical parts of a given 2qcfa
may be probabilistic as well.
For each input $x$, we may define a probability $p_{acc}(x)$ that a given
2qcfa $M$ eventually enters a classical accepting state, and a probability
$p_{rej}(x)$ that $M$ eventually enters a rejecting state.
A given computation is assumed to halt when either an accepting or rejecting
classical state is reached, so the above events are mutually exclusive.
We say that a given machine $M$ recognizes a language $L\subseteq\Sigma^{\ast}$
with one-sided error $\epsilon$ if for all $x\in\Sigma^{\ast}$ we have
$p_{acc}(x)+p_{rej}(x)=1$, $p_{acc}(x)=1$ if $x\in L$, and
$p_{rej}(x)\geq 1-\epsilon$ if $x\not\in L$.
Other notions of recognition such as two-sided error, zero error, etc., may
be defined analogously, but we will only consider one-sided error in this
paper.

A natural extension of our model is to allow POVM-type measurements
(see~\cite{Preskill98}, for instance) rather than orthogonal measurements. 
In fact this does not change the power of the model since POVM-type
measurements may be simulated by orthogonal measurements and unitary
operators on (possibly) larger quantum systems.
It may be the case that one may reduce the number of states required for
various tasks using POVMs, although it is questionable whether this has any
physical significance.


\section{Recognizing Palindromes}
\label{sec:palindromes}

In this section we prove that 2qcfa's can recognize palindromes with any fixed
error bound $\epsilon>0$, which is an impossible task for classical
probabilistic 2-way finite automata.
We first define a 2qcfa for this language that uses a quantum register
having three orthogonal states, since this is easier to describe than the two
orthogonal state (i.e., single qubit) case.
Once we have this, it is simple to modify the 2qcfa so that it requires a
single qubit register, due to the fact that the three orthogonal state machine
uses only real amplitudes, along with the fact that there is a natural mapping
from the unit sphere in real three-dimensional Euclidean space to the unit
sphere in a two-dimensional complex Hilbert space.

\begin{theorem}
\label{thm:palindromes}
For any $\epsilon>0$ there exists a 2qcfa $M$ operating as follows.
For any input $x\in\{a,b\}^{\ast}$, if $x$ is a palindrome then $M$ accepts
$x$ with certainty, and if $x$ is not a palindrome then $M$ accepts $x$ with
probability at most $\epsilon$ and rejects $x$ otherwise.
\end{theorem}

\noindent
In order to prove Theorem~\ref{thm:palindromes}, we consider the
$3\times 3$ integer matrices $A$ and $B$, defined as follows.
\begin{equation}
A = \left(\!\!
\begin{array}{rrr}
4  & 3 & 0\\
-3 & 4 & 0\\
0  & 0 & 5
\end{array}
\right),
\;\;\;\;\;
B = \left(\!\!
\begin{array}{rrr}
4  & 0 & 3\\
0  & 5 & 0\\
-3 & 0 & 4
\end{array}
\right).
\label{eq:A_and_B}
\end{equation}
Also define a function $f:\Zed^{3}\rightarrow\Zed$ as
\[
f(u) = 4u[1]+3u[2]+3u[3],
\]
for each $u\in\Zed^{3}$, and define a set $K\subseteq\Zed^{3}$ as
\[
K = \left\{u\in\Zed^{3}:
u[1]\not\equiv 0\,(\bmod\: 5),\;
f(u)\not\equiv 0\,(\bmod\: 5),\,\mbox{and}\;
u[2]\cdot u[3]\equiv 0\,(\bmod\: 5)
\right\}.
\]

\begin{lemma}
If $u\in K$, then $Au\in K$ and $Bu\in K$.
\label{lemma:K_closure}
\end{lemma}
{\bf Proof.}
We show that $u\in K$ implies $Au\in K$; the proof for $Bu\in K$ is similar.
Write $u = (a,\,b,\,c)^{T}$, so that $Au = (4a + 3b,\, -3a + 4b,\, 5c)^{T}$.
We immediately see $(Au)[2]\cdot(Au)[3]\equiv 0\:(\bmod\,5)$, so it remains to
show $(Au)[1]\not\equiv 0\:(\bmod\,5)$ and $f(Au)\not\equiv 0\:(\bmod\,5)$.
Since $u\in K$, we have
\begin{eqnarray}
&a\:\not\equiv\: 0 \pmod{5}&\label{eq:cong1}\\
&f(u) \:=\: 4a + 3 b + 3c \:\not\equiv\: 0\pmod{5}&\label{eq:cong2}
\end{eqnarray}
and either $b\equiv 0\,(\bmod\: 5)$ or $c\equiv 0\,(\bmod\: 5)$.

Suppose first that $b\equiv 0\,(\bmod\: 5)$.
Then we have $(Au)[1]\equiv 4a\,(\bmod\:5)$ and
\[
f(Au) \:=\: 4(4a+3b) + 3(-3a + 4b) + 3(5c)\:\equiv\: 2a\,(\bmod\:5).
\]
Thus $(Au)[1]\not\equiv 0\,(\bmod\:5)$ and $f(Au)\not\equiv 0\,(\bmod\:5)$ by
(\ref{eq:cong1}).

Now suppose  $c\equiv 0\,(\bmod\: 5)$.
Then
\begin{eqnarray*}
(Au)[1] & = & 4a + 3b\\
& \equiv & 4a + 3b + 3c \pmod{5}\\
& \equiv & f(u)\pmod{5}
\end{eqnarray*}
and
\begin{eqnarray*}
f(Au) & = & 4(4a + 3b) + 3(-3a+4b) + 3(5c)\\
& \equiv & 2a + 4b \pmod{5}\\
& \equiv & 3(4a + 3b + 3c) \pmod{5}\\
& \equiv & 3f(u) \pmod{5},
\end{eqnarray*}
so that $(Au)[1]\not\equiv 0\,(\bmod\:5)$ and $f(Au)\not\equiv 0\,(\bmod\:5)$
by (\ref{eq:cong2}), which completes the proof.
\qed

\begin{lemma}
Let $u\in\Zed^{3}$ satisfy $u=Av=Bw$ for $v,w\in\Zed^{3}$.
Then $u\not\in K$.
\label{lemma:not_in_K}
\end{lemma}
{\bf Proof.}
Assume $u = Av = Bw$ for $u,v,w \in\Zed^{3}$, so that
$A^{-1}u,\,B^{-1}u\in\Zed^{3}$.
Since $(B^{-1}u)[2]\in\Zed$ we conclude $u[2]\equiv 0\:(\bmod\:5)$, and since
$(A^{-1}u)[1]\in\Zed$ we conclude $4u[1] - 3u[2]\equiv 0\:(\bmod\:25)$.
Together these congruences imply $u[1]\equiv 0\:(\bmod\:5)$, and hence
$u\not\in K$.
\qed

\begin{lemma}
\label{lemma:pal}
Let
\[
u = Y_{1}^{-1}\,\cdots\,Y_{n}^{-1}\,X_{n}\,\cdots\,X_{1}(1,0,0)^{T},
\]
where $X_{j},Y_{j}\in\{A,B\}$.
If $X_{j} = Y_{j}$ for $1\leq j\leq n$, then $u[2]^{2} + u[3]^{2} = 0$.
Otherwise, $u[2]^{2} + u[3]^{2} > 25^{-n}$.
\end{lemma}
{\bf Proof.}
If $X_{j} = Y_{j}$ for $1\leq j\leq n$, then we clearly have $u = (1,0,0)^{T}$,
and thus $u[2]^{2} + u[3]^{2} = 0$.

Next suppose there exists $j$ such that $X_{j}\not=Y_{j}$.
Note that $\|u\| = 1$, since $5^{-1}X_{j}$ and $5Y_{j}^{-1}$ are unitary for
each $j$, and further note that $25^{n}u$ is integer valued.
To prove the lemma it therefore suffices to prove $u\not=\pm(1,0,0)^{T}$, since
$|u[1]|<1$ implies $|u[1]|\leq 1 - 25^{-n}$, and therefore
\[
u[2]^{2} + u[3]^{2} = 1 - u[1]^{2} \geq 1 - (1-25^{-n})^2 > 25^{-n}.
\]
Let $k$ be the largest index such that $X_{k}\not=Y_{k}$, and without loss of
generality suppose $X_{k}=A$, $Y_{k}=B$.
Write $v = X_{k-1}\cdots X_{1}(1,0,0)^{T}$ and
$w = Y_{k-1}\cdots Y_{1}(1,0,0)^{T}$.
Since $(1,0,0)^{T}\in K$, we must have $Av,Bw\in K$ by
Lemma~\ref{lemma:K_closure}.
By Lemma~\ref{lemma:not_in_K} this implies $Av\not=Bw$, since
$Av=Bw$ contradicts the fact that $Av,Bw\in K$.
Since $X_{j} = Y_{j}$ for $j>k$, we therefore have
$Y_{n}\cdots Y_{1}(1,0,0)^{T}\not=X_{n}\cdots X_{1}(1,0,0)^{T}$ and thus
$u = Y_{1}^{-1}\,\cdots\,Y_{n}^{-1}\,X_{n}\,\cdots\,X_{1}(1,0,0)^{T} \not=
(1,0,0)^{T}$.
By similar reasoning, $u\not=(-1,0,0)^{T}$ since
$(-1,0,0)^{T}\in K$ and hence
$Y_{n}\cdots Y_{1}(-1,0,0)^{T}\not=X_{n}\cdots X_{1}(1,0,0)^{T}$.
\qed

\noindent {\bf Proof of Theorem~\ref{thm:palindromes}.}
Define $U_{a}$ and $U_{b}$ to be unitary operators as follows:
\begin{alignat*}{2}
U_{a}\ket{q_{0}} &= \frac{4}{5}\ket{q_{0}} - \frac{3}{5}\ket{q_{1}},\qquad
& U_{b}\ket{q_{0}} &= \frac{4}{5}\ket{q_{0}} - \frac{3}{5}\ket{q_{2}},\\
U_{a}\ket{q_{1}} &= \frac{3}{5}\ket{q_{0}} + \frac{4}{5}\ket{q_{1}},\qquad
& U_{b}\ket{q_{1}} &= \ket{q_{1}},\\
U_{a}\ket{q_{2}} &= \ket{q_{2}},\qquad
& U_{b}\ket{q_{2}} &= \frac{3}{5}\ket{q_{0}} + \frac{4}{5}\ket{q_{2}},
\end{alignat*}
and define $M$ to be a 2qcfa as described in Figure~\ref{fig:palindromes}.
\begin{figure}[ht]
\begin{flushleft}
\hrulefill\\[3mm]
\inpar{0mm}{1mm}{Repeat ad infinitum:}
\inpar{7.5mm}{1mm}{Move the tape head to the first input symbol and set the
        quantum state to $\ket{q_{0}}$.}
\inpar{7.5mm}{1mm}{While the currently scanned symbol is not \$, do the
        following: \hfill (I)}
\inpar{15mm}{1mm}{Perform $U_{\sigma}$ on the quantum state, for $\sigma$
        denoting the currently scanned symbol.}
\inpar{15mm}{1mm}{Move the tape head one square to the right.}
\inpar{7.5mm}{1mm}{Move the tape head left until the \cent\ symbol is reached.}
\inpar{7.5mm}{1mm}{Move the tape head one square to the right.}
\inpar{7.5mm}{1mm}{While the currently scanned symbol is not \$, do the
        following: \hfill (II)}
\inpar{15mm}{1mm}{Perform $U_{\sigma}^{-1}$ on the quantum state, for $\sigma$
        denoting the currently scanned symbol.}
\inpar{15mm}{1mm}{Move the tape head one square to the right.}
\inpar{7.5mm}{1mm}{Measure the quantum state: if the result is not $q_{0}$ then
        \underline{reject}.}
\inpar{7.5mm}{1mm}{Set $b=0$.}
\inpar{7.5mm}{1mm}{While the currently scanned symbol is not \cent, do the
        following: \hfill (III)}
\inpar{15mm}{1mm}{Simulate $k$ coin-flips. Set $b=1$ in case all results
        are not ``heads''.}
\inpar{15mm}{1mm}{Move the tape head one square to the left.}
\inpar{7.5mm}{1mm}{If $b=0$, \underline{accept}.}
\hrulefill
\end{flushleft}
\caption{A 2qcfa for palindromes.}
\label{fig:palindromes}
\end{figure}
The parameter $k$ will be specified below according to the error bound
$\epsilon$.

The action of $M$ on input $x = x_{1}x_{2}\cdots x_{n}$ is as follows.
The machine starts with its quantum state in superposition $\ket{q_{0}}$.
As while-loop (I) is executed, the tape head moves over each input symbol and
performs either the transformation $U_{a}$ or $U_{b}$ on the quantum state
(depending on whether the symbol scanned is $a$ or $b$).
Letting $X_{j}$ denote the matrix $A$ or $B$, as defined in (\ref{eq:A_and_B}),
depending on whether $x_{j}$ is $a$ or $b$, we see that the superposition of
the quantum state of $M$ after performing loop (I) is
\[
\alpha_{0}\ket{q_{0}} + \alpha_{1}\ket{q_{1}} + \alpha_{2}\ket{q_{2}}
\]
for
$(\alpha_{0},\alpha_{1},\alpha_{2})^{T}=5^{-n}X_{n}\cdots X_{1}(1,0,0)^{T}$.
At this point, the tape head is moved back to the first input symbol and
while-loop (II) is performed.
A process similar to while-loop (I) is performed (except the inverses of
$U_{a}$ and $U_{b}$ are applied instead of $U_{a}$ and $U_{b}$), yielding
superposition
\[
\beta_{0}\ket{q_{0}} + \beta_{1}\ket{q_{1}} + \beta_{2}\ket{q_{2}}
\]
for $(\beta_{0},\beta_{1},\beta_{2})^{T}=
X^{-1}_{n}\cdots X^{-1}_{1}X_{n}\cdots X_{1}(1,0,0)^{T}$.
Now the quantum state is measured: $M$ rejects with probability
$p_{rej} = \beta_{1}^{2} + \beta_{2}^{2}$, and otherwise the quantum state
collapses to $\ket{q_{0}}$ with probability $\beta_{0}^{2}$.
By Lemma~\ref{lemma:pal} we conclude $p_{rej} = 0$ in case $x$ is a
palindrome, and $p_{rej}> 25^{-n}$ otherwise.
Finally, $M$ sets the variable $b$ (stored in its classical internal state) to
$0$, executes while-loop (III), and accepts if the while-loop terminates with
$b$ still set to $0$; it may be checked that this happens with probability
$p_{acc} = 2^{-k(n+1)}$.

This sequence of steps is repeated indefinitely, causing $M$ to eventually
reject with probability
\[
\sum_{j\geq 0}(1-p_{acc})^{j}(1-p_{rej})^{j}p_{rej} \:=\:
\frac{p_{rej}}{p_{acc} + p_{rej} - p_{acc}p_{rej}}
\]
and accept with probability
\[
\sum_{j\geq 0}(1-p_{acc})^{j}(1-p_{rej})^{j+1}p_{acc} \:=\:
\frac{p_{acc} - p_{acc}p_{rej}}{p_{acc} + p_{rej} - p_{acc}p_{rej}}.
\]
These probabilities clearly sum to 1, and the probability of acceptance is
therefore 1 in case $x$ is a palindrome.
Letting $k\geq \max\{\log 25, -\log\epsilon\}$, we see that
if $x$ is not a palindrome, then $M$ rejects with probability at least
$1 - \epsilon$, which completes the proof.
\qed

We now outline how this 2qcfa may be modified so that a only single qubit is
used.
Define a mapping $\Phi$ from the unit sphere in $\mathbb{R}^3$ to the
unit sphere in $\mathbb{C}^2$ as follows:
\[
\Phi(\cos\phi\ket{q_0}+\sin\phi\,\sin\psi\ket{q_1}+\sin\phi\,\cos\psi\ket{q_2})
\:=\: e^{-i\psi/2}\cos\frac{\phi}{2}\ket{0}+e^{i\psi/2}\sin\frac{\phi}{2}
\ket{1},
\]
and define
\begin{alignat*}{2}
\widehat{U}_a\ket{0} &= 
\phantom{-}\cos\frac{\theta}{2}\ket{0}-i\sin\frac{\theta}{2}\ket{1},\qquad &
\widehat{U}_b\ket{0} &=
\phantom{-}\cos\frac{\theta}{2}\ket{0}+\sin\frac{\theta}{2}\ket{1},\\
\widehat{U}_a\ket{1} &= 
-i\sin\frac{\theta}{2}\ket{0}+\cos\frac{\theta}{2}\ket{1},\qquad &
\widehat{U}_b\ket{1} &= 
-\sin\frac{\theta}{2}\ket{0}+\cos\frac{\theta}{2}\ket{1},
\end{alignat*}
for $\theta = \tan^{-1}(4/3)$.
It may be verified that the following relations hold:
\begin{eqnarray*}
\widehat{U}_a\,\Phi(\alpha_0\ket{q_0}+\alpha_1\ket{q_1}+\alpha_2\ket{q_2}) &=&
e^{i\phi}\Phi(U_a(\alpha_0\ket{q_0}+\alpha_1\ket{q_1}+\alpha_2\ket{q_2}))\\
\widehat{U}_b\,\Phi(\alpha_0\ket{q_0}+\alpha_1\ket{q_1}+\alpha_2\ket{q_2}) &=&
e^{i\phi}\Phi(U_b(\alpha_0\ket{q_0}+\alpha_1\ket{q_1}+\alpha_2\ket{q_2})),
\end{eqnarray*}
where $e^{i\phi}$ represents a phase factor (possibly depending on
$\alpha_0$, $\alpha_1$, and $\alpha_2$) that will not affect the operation
of the machine.
The proof of this claim follows from a much more general relationship between
rigid rotations in three dimensions and unitary transformations in two
dimensions; see, for instance,~\cite{MathewsW70} for further discussion.
(See also Section~2.3.2 in~\cite{Preskill98}.)
Note here that we have exchanged the $x$ and $z$ coordinates from the mappings
described in these references in order to allow the observations to function
correctly.
Clearly we have that an observation of the state $\Phi(\ket{q_0})$ (in the
$\{\ket{0},\ket{1}\}$ basis) yields $\ket{0}$ with probability 1, and an
observation of $\Phi(\alpha_0\ket{q_0}+\alpha_1\ket{q_1}+\alpha_2\ket{q_2})$
yields $\ket{1}$ with probability at least $(1-\sqrt{1-\delta})/2\geq\delta/4$
in case $\alpha_1^2+\alpha_2^2\geq \delta$.
Thus, by substituting transformation $\widehat{U}_a$ for $U_a$,
$\widehat{U}_b$ for $U_b$, adjusting $k$ as necessary, and letting $\ket{0}$
be the initial state of the quantum register in the machine constructed above,
we obtain a 2qcfa for palindromes that uses a single qubit.


\section{Recognizing $a^n b^n$}
\label{sec:anbn}

The second language that we consider is $\{a^n b^n| n\in\Natural\}$.
It is non-regular but can be recognized by a 2-way 
probabilistic finite automaton \cite{Freivalds81}. 
However, any 2-way probabilistic automaton recognizing
it runs in exponential expected time \cite{GreenbergW86}.
(More generally, a similar result is true for 2-way probabilistic
automata recognizing any nonregular language \cite{DworkS90,KanepsF91}.)

In the quantum world, this language can be recognized by
a 2qfa \cite{KondacsW97}. However, this 2qfa uses superpositions
where the head of the qfa is in different places for different
components of the superposition and, therefore, cannot be
implemented with a quantum part of finite size.
In this paper, we show that this language can be also
recognized by a 2qcfa in polynomial time.
This result is proved by a method similar to Theorems~8 and 14 of
\cite{AmbainisF98}.

\begin{theorem}
For any $\epsilon>0$, there is a 2qcfa $M$ that accepts 
any $x\in\{a^n b^n |n\in\Natural\}$ with certainty,
rejects $x\notin\{a^n b^n|n\in\Natural\}$ with probability
at least $1-\epsilon$ and halts in expected time $O(m^4)$
where $m$ is the length of the word $x$.
\end{theorem}

\noindent
{\bf Proof:}
The main idea is as follows: 

We consider a qcfa $M$ with 2 quantum states $\ket{q_0}$ and $\ket{q_1}$.
$M$ starts in the state $\ket{q_0}$.
Every time when $M$ reads $a$, the state is rotated
by angle $\alpha=\sqrt{2}\pi$ and every time when $M$ reads $b$, the state
is rotated by $-\alpha$. When the end of the word is reached,
$M$ measures the state. If it is $\ket{q_1}$, the word is rejected.
Otherwise, the whole process is repeated.

If the number of $a$'s is
equal to the number of $b$'s, rotations cancel one another
and the final state is $q_0$. 
Otherwise, the final state is different from $q_0$
because $\sqrt{2}$ is irrational.
Moreover, the amplitude of $q_1$ in the final state
is sufficiently large\footnote{This relies on a property
of $\sqrt{2}$ and is not true for an arbitrary irrational
number instead of $\sqrt{2}$.}.
Therefore, repeating the above process $O(n^2)$ times
guarantees getting $q_1$ at least once (and rejecting the input)
with a high probability.

We also need that $M$ halts on inputs $x\in\{a^n b^n|n\in\Natural\}$
(instead of repeating the above process forever).
To achieve that, we periodically execute a subroutine that accepts
with a small probability $\frac{c}{n^2}$. 
If the word is not in language, this does not have much influence
because this probability is much smaller than the probability
of getting $q_1$ in one run.
The resulting automaton is described in Figure \ref{fig:anbn}.

\begin{figure}[!ht]
\begin{flushleft}
\hrulefill\\[3mm]
\inpar{0mm}{1mm}{Check (classically) whether the input is of form 
$a^*b^*$. If not, \underline{reject}.}
\inpar{0mm}{1mm}{Otherwise, repeat ad infinitum:}
\inpar{7.5mm}{1mm}{Move the tape head to the first input symbol and set the
        quantum state to $\ket{q_{0}}$.}
\inpar{7.5mm}{1mm}{While the currently scanned symbol is not \$, do the
        following: \hfill (I)}
\inpar{15mm}{1mm}{If the currently scanned symbol is $a$, perform 
$U_{\alpha}$ on the quantum state.}
\inpar{15mm}{1mm}{If the currently scanned symbol is $b$, perform 
$U_{-\alpha}$ on the quantum state.}
\inpar{15mm}{1mm}{Move the tape head one square to the right.}
\inpar{7.5mm}{1mm}{Measure the quantum state. If the result is
not $q_0$, \underline{reject}.}
\inpar{7.5mm}{1mm}{Two times repeat:\hfill (II)}
\inpar{15mm}{1mm}{Move the tape head to the first input symbol}
\inpar{15mm}{1mm}{Move the tape head one square to the right.}
\inpar{15mm}{1mm}{While the currently scanned symbol is not 
$\cent$ or \$, do the following:\hfill (III)}
\inpar{22.5mm}{1mm}{Simulate a coin flip. If the result is "heads",
move right. Otherwise, move left.}  
\inpar{7.5mm}{1mm}{If both times the process ends at the right end-marker \$, do:}
\inpar{15mm}{1mm}{Simulate $k$ coin-flips. If all results
        are not ``heads'', \underline{accept}.}
\hrulefill
\end{flushleft}
\caption{A 2qcfa for $a^n b^n$.}
\label{fig:anbn}
\end{figure}

Next, we show that this automaton recognizes $\{a^n b^n|n\in\Natural\}$.
It is enough to consider its action on words of form $a^n b^{n'}$
(because all other words are rejected by it at the very beginning).
We start with two lemmas that bound the probabilities of
accepting after loop (I) and rejecting after loop (II).

\begin{lemma}
If the input is $x=a^n b^{n'}$ and $n'\neq n$,
$M$ rejects after loop (I) with probability at least 
$1/2(n-n')^2$. 
\end{lemma}

\noindent
{\bf Proof:}
In this case, the state $\ket{q_0}$ gets rotated by 
$\sqrt{2} (n-n') \pi$. 
The superposition after rotating $\ket{q_0}$ by $\sqrt{2}(n-n')\pi$ is
\[ \cos(\sqrt{2}(n-n')\pi) \ket{q_0} + \sin(\sqrt{2}(n-n')\pi) \ket{q_1} .\]
The probability of observing $\ket{q_1}$ is $\sin^2(\sqrt{2}(n-n')\pi)$.
We bound this probability from below.

Let $k$ be the closest integer
to $\sqrt{2} (n-n')$. Assume that $\sqrt{2} (n-n')>k$.
(The other case is symmetric.)
Then, $k\leq \sqrt{2(n-n')^2-1}$ (because $k^2$ is integer
and $2(n-n')^2-1$ is the largest integer that is smaller
than $(\sqrt{2}(n-n'))^2$). We have 
\[ (\sqrt{2}(n-n')-\sqrt{2(n-n')^2-1})(\sqrt{2}(n-n')+
\sqrt{2(n-n')^2-1})=2(n-n')^2-2(n-n')^2+1=1 ,\]
\[ \sqrt{2}(n-n')-k\geq \sqrt{2}(n-n')-\sqrt{2(n-n')^2-1}=
\frac{1}{\sqrt{2}(n-n')+\sqrt{2(n-n')^2-1}}>
\frac{1}{2\sqrt{2}(n-n')} .\]

We have $0<\sqrt{2}(n-n')-k<1/2$ (because $k$ is the closest integer).
For any $x\in[0, 1/2]$, $\sin(x\pi)\geq 2x$
(because this is true for $x=0$ and $x=1/2$ and $\sin$ is concave in this
interval). Therefore, 
\[ \sin^2(\sqrt{2}(n-n')\pi)=\sin^2((\sqrt{2}(n-n')-k)\pi)\geq 
4 (\sqrt{2}(n-n')-k)^2\geq 4 \left(\frac{1}{2\sqrt{2}(n-n')}\right)^2
= \frac{1}{2 (n-n')^2}.\] 
\qed

\begin{lemma}
Each execution of (II) leads to acceptance with probability 
$1/2^k(n+n'+1)^2$.
\end{lemma}

\noindent
{\bf Proof:}
Each loop (III) is just a random walk starting at location 1 
(the first symbol of $a^n b^{n'}$) and ending either at location 0
(the left end-marker $\cent$) or location $n+n'+1$ (the right end-marker \$).
It is a standard result in probability theory (see Chapter 14.2 of 
\cite{Feller67}) that the probability of reaching the location $n+n'+1$
is exactly $\frac{1}{n+n'+1}$.
Repeating it twice and flipping $k$ coins afterwards
gives the probability $1/2^k(n+n'+1)^2$.
\qed

We select $k=1+\lceil\log \epsilon\rceil$.
If $x=a^n b^n$, then the loop (I) always returns $\ket{q_0}$
to $\ket{q_0}$ and $M$ never rejects.
The probability of $M$ accepting after $c n^2$ executions of (II) is
\[ 1-\left(1-\frac{1}{2^k(n+n'+1)^2}\right)^{c n^2}\]
and this can be made arbitrarily close to 1 by selecting an appropriate 
constant $c$.

On the other hand, if $x=a^n b^{n'}$ and $n\neq n'$,
$M$ rejects after (I) with probability $p_{rej}>1/2(n-n')^2$ and
accepts after (II) with probability $p_{acc} 1/2^k(n+n'+1)^2\leq 
\epsilon/2(n+n'+1)^2$.
If this is repeated indefinitely, the probability of rejecting is
\[
\sum_{k\geq 0}(1-p_{acc})^{k}(1-p_{rej})^{k}p_{rej} \:=\:
\frac{p_{rej}}{p_{acc} + p_{rej} - p_{acc}p_{rej}} > 
\frac{p_{rej}}{p_{acc} + p_{rej}} > \frac{1/2}{1/2+\epsilon/2}=
\frac{1}{1+\epsilon}>1-\epsilon .\]

In both cases, the expected number of iterations of (I) and (II) 
before $M$ accepts or rejects is $O((n+n')^2)$ (because, in every
iteration, $M$ accepts or rejects with probability at least $c/(n+n')^2$).
Loop (I) takes $O(n+n')$ time and each random walk in (II) takes 
$O((n+n')^2)$ time. Hence, the expected running time of $M$ is
at most $O((n+n')^4)$.  
\qed


\section{Conclusion}
\label{sec:conclusion}

In this paper we have introduced 2-way finite automata with quantum and
classical states, and given two examples of languages for which 2qcfa's
outperform classical probabilistic 2-way finite automata: 
$L_{pal}=\{x\in\{a,b\}^{\ast}\,|\,x = x^R\}$ 
and $L_{eq}=\{a^nb^n\,|\,n\in\Natural\}$.
It is natural to ask what other languages can be recognized by 2qcfa's.
For instance, can any of the following languages be recognized by 2qcfa's?
\begin{mylist}{\parindent}
\item $L_{\mathrm{middle}}=\left\{xay\,|\,x,y\in\{a,b\}^{\ast},\:
|x|=|y|\right\}$.
\item $L_{\mathrm{balanced}} = \left\{x\in\{\,(\,,\,)\,\}^{\ast}\,|\,
\mbox{parentheses in $x$ are balanced}\right\}$.
\item $L_{\mathrm{square}}=\left\{a^nb^{n^2}\,|\,n\in\mathbb{N}\right\}$.
\item $L_{\mathrm{power}}=\left\{a^nb^{2^n}\,|\,n\in\mathbb{N}\right\}$.
\end{mylist}
If so, can any of these languages be recognized by polynomial time 2qcfa's?


\bibliographystyle{plain}

\begin{thebibliography}{10}

\bibitem{AharonovK+98}
D.~Aharonov, A.~Kitaev, and N.~Nisan.
\newblock Quantum circuits with mixed states.
\newblock In {\em Proceedings of the Thirtieth Annual ACM Symposium on Theory
  of Computing}, pages 20--30, 1998.

\bibitem{AmbainisF98}
A.~Ambainis and R.~Freivalds.
\newblock 1-way quantum finite automata: strengths, weaknesses and
  generalizations.
\newblock In {\em Proceedings of the 39th Annual Symposium on Foundations of
  Computer Science}, pages 332--341, 1998.

\bibitem{AmbainisN+99}
A.~Ambainis, A.~Nayak, A.~Ta-Shma, and U.~Vazirani.
\newblock Dense quantum coding and a lower bound for 1-way quantum automata.
\newblock In {\em Proceedings of the Thirty-First Annual ACM Symposium on
  Theory of Computing}, pages 376--383, 1999.

\bibitem{Berthiaume97}
A.~Berthiaume.
\newblock Quantum computation.
\newblock In L.~Hemaspaandra and A.~Selman, editors, {\em Complexity Theory
  Retrospective II}, pages 23--50. Springer, 1997.

\bibitem{BrodskyP99}
A.~Brodsky and N.~Pippenger.
\newblock Characterizations of 1-way quantum finite automata.
\newblock Los Alamos Preprint Archive, quant-ph/9903014, 1999.

\bibitem{DworkS90}
C.~Dwork and L.~Stockmeyer.
\newblock A time-complexity gap for two-way probabilistic finite state
  automata.
\newblock {\em SIAM Journal of Computing}, 19:1011--1023, 1990.

\bibitem{DworkS92}
C.~Dwork and L.~Stockmeyer.
\newblock Finite state verifiers {I}: the power of interaction.
\newblock {\em Journal of the ACM}, 39(4):800--828, 1992.

\bibitem{Feller67}
W.~Feller.
\newblock {\em An Introduction to Probability Theory and Its Applications},
  volume~I.
\newblock Wiley, 1967.

\bibitem{Freivalds81}
R.~Freivalds.
\newblock Probabilistic two-way machines.
\newblock In {\em Proceedings of the International Symposium on Mathematical
  Foundations of Computer Science}, volume 188 of {\em Lecture Notes in
  Computer Science}, pages 33--45. Springer-Verlag, 1981.

\bibitem{GreenbergW86}
A.~Greenberg and A.~Weiss.
\newblock A lower bound for probabilistic algorithms for finite state machines.
\newblock {\em Journal of Computer and System Sciences}, 33(1):88--105, 1986.

\bibitem{Grover96}
L.~Grover.
\newblock A fast quantum mechanical algorithm for database search.
\newblock In {\em Proceedings of the Twenty-Eighth Annual ACM Symposium on
  Theory of Computing}, pages 212--219, 1996.

\bibitem{Holevo73}
A.~Holevo.
\newblock Bounds for the quantity of information transmitted by a quantum
  communication channel.
\newblock {\em Problemy Peredachi Informatsii}, 9(3):3--11, 1973.
\newblock English translation in {\em Problems of Information Transmission}
  {\bf 9}, 1973.

\bibitem{KanepsF91}
J.~Ka\c{n}eps and R.~Freivalds.
\newblock Running time to recognize nonregular languages by 2-way probabilistic
  automata.
\newblock In {\em Proceedings of the 18th International Colloquium on Automata,
  Languages and Programming}, volume 510 of {\em Lecture Notes in Computer
  Science}, pages 174--185, 1991.

\bibitem{Kitaev97}
A.~Kitaev.
\newblock Quantum computations: algorithms and error correction.
\newblock {\em Russian Mathematical Surveys}, 52(6):1191--1249, 1997.

\bibitem{KondacsW97}
A.~Kondacs and J.~Watrous.
\newblock On the power of quantum finite state automata.
\newblock In {\em Proceedings of the 38th Annual Symposium on Foundations of
  Computer Science}, pages 66--75, 1997.

\bibitem{MathewsW70}
J.~Mathews and R.~Walker.
\newblock {\em Mathematical Methods of Physics}.
\newblock W.~A.~Benjamin, Inc., New York, second edition, 1970.

\bibitem{MooreC97}
C.~Moore and J.~Crutchfield.
\newblock Quantum automata and quantum grammars.
\newblock Technical Report Working Paper 97-07-062, Santa Fe Institute, 1997.
\newblock To appear in {\em Theoretical Computer Science}.

\bibitem{Nayak99}
A.~Nayak.
\newblock Optimal lower bounds for quantum automata and random access codes.
\newblock Los Alamos Preprint Archive, quant-ph/9904093, 1999.

\bibitem{Preskill98}
J.~Preskill.
\newblock Lecture notes for physics 229: Quantum information and computation.
\newblock California Institute of Technology. Available at
http://theory.caltech.edu/people/preskill/ph229, 1998.

\bibitem{Shor97}
P.~Shor.
\newblock Polynomial-time algorithms for prime factorization and discrete
  logarithms on a quantum computer.
\newblock {\em SIAM Journal on Computing}, 26(5):1484--1509, 1997.

\end{thebibliography}

\end{document}